\documentstyle[12pt,epsfig]{article}
\font\tenrm=cmr10

 1
\font\elevenrm=cmr10 scaled\magstep 1
 1

\begin{document}
\begin{titlepage}
\def\baselinestretch{1.2}


\vspace*{\fill}
\begin{center}
{\large
{\bf   Physics of Self-Interacting Electroweak Bosons}}
\footnote{ Plenary Talk given  at the ``Beyond the Standard Model IV",
12-16 Dec. 1994, Granlibakken, Lake Tahoe, CA, USA.}

\vspace*{0.5cm}

Fawzi Boudjema \\

{\it Laboratoire de Physique Th\'eorique
EN{\large S}{\Large L}{\large A}PP}
\footnote{ URA 14-36 du CNRS, associ\'ee \`a l'E.N.S de Lyon
et \`a l'Universit\'e de Savoie.}\\
{\it B.P.110, 74941 Annecy-Le-Vieux Cedex, France} \\
{\tenrm E-mail:BOUDJEMA@LAPPHP8.IN2P3.FR}

\end{center}
\vspace*{\fill}

\centerline{ {\bf Abstract} }
\baselineskip=14pt
\noindent
 {\small  I will argue why and how it is that precise measurements of the
self-couplings
 of the weak vector bosons are a vista on the mechanism of symmetry breaking.
Guided by what we have learnt from the present precision data, it is suggested
which of the many so-called anomalous self-couplings should be given priority
in
future searches. Expected limits from the upcoming colliders
on the parameters describing non minimal couplings
are updated. I will also point at the complementarity between the LHC and the
Next Linear Collider as concerns $W$ physics and discuss some of the important
issues about radiative corrections and backgrounds that need further studies in
order that one conducts high precision analysis at high energies.}
\vspace*{\fill}


\vspace*{0.1cm}
\rightline{ENSLAPP-A-513/95}
\rightline{hep-ph/9504409}
\rightline{March 1995}
\end{titlepage}
\baselineskip=18pt


\newcommand{\beq}{\begin{equation}}
\newcommand{\eeq}{\end{equation}}

\newcommand{\beqn}{\begin{eqnarray}}
\newcommand{\eeqn}{\end{eqnarray}}

\newcommand{\ra}{\rightarrow}

\newcommand{\su}{$ SU(2) \times U(1)\,$}

\newcommand{\gag}{$\gamma \gamma$ }
\newcommand{\gam}{\gamma \gamma }

\newcommand{\np}{Nucl.\,Phys.\,}
\newcommand{\pl}{Phys.\,Lett.\,}
\newcommand{\pr}{Phys.\,Rev.\,}
\newcommand{\prl}{Phys.\,Rev.\,Lett.\,}
\newcommand{\prep}{Phys.\,Rep.\,}
\newcommand{\zp}{Z.\,Phys.\,}
\newcommand{\sovjnp}{{\em Sov.\ J.\ Nucl.\ Phys.\ }}
\newcommand{\nuclinst}{{\em Nucl.\ Instrum.\ Meth.\ }}
\newcommand{\annp}{{\em Ann.\ Phys.\ }}
\newcommand{\intjmp}{{\em Int.\ J.\ of Mod.\  Phys.\ }}

\newcommand{\eps}{\epsilon}
\newcommand{\mw}{M_{W}}
\newcommand{\mww}{M_{W}^{2}}
\newcommand{\mwmw}{M_{W}^{2}}
\newcommand{\mhmh}{M_{H}^2}
\newcommand{\mz}{M_{Z}}
\newcommand{\mzz}{M_{Z}^{2}}

\newcommand{\lra}{\leftrightarrow}
\newcommand{\tr}{{\rm Tr}}

\newcommand{\dkg}{\Delta \kappa_{\gamma}}
\newcommand{\dkz}{\Delta \kappa_{Z}}
\newcommand{\dz}{\delta_{Z}}
\newcommand{\dgz}{\Delta g^{1}_{Z}}
\newcommand{\dgzt}{$\Delta g^{1}_{Z}\;$}
\newcommand{\la}{\lambda}
\newcommand{\lag}{\lambda_{\gamma}}
\newcommand{\laz}{\lambda_{Z}}
\newcommand{\lnl}{L_{9L}}
\newcommand{\lnr}{L_{9R}}
\newcommand{\lt}{L_{10}}
\newcommand{\lu}{L_{1}}
\newcommand{\ld}{L_{2}}

\newcommand{\epm}{$e^{+} e^{-}\;$}
\newcommand{\epemt}{$e^{+} e^{-}\;$}
\newcommand{\epem}{e^{+} e^{-}\;}
\newcommand{\eeww}{e^{+} e^{-} \ra W^+ W^- \;}
\newcommand{\eewwt}{$e^{+} e^{-} \ra W^+ W^- \;$}
\newcommand{\ppwg}{p p \ra W \gamma}
\newcommand{\ppwz}{pp \ra W Z}
\newcommand{\ppwgt}{$p p \ra W \gamma \;$}
\newcommand{\ppwzt}{$pp \ra W Z \;$}
\newcommand{\gamgamt}{$\gamma \gamma \;$}
\newcommand{\gamgam}{\gamma \gamma \;}
\newcommand{\egamt}{$e \gamma \;$}
\newcommand{\egam}{e \gamma \;}
\newcommand{\gamgamtwwt}{$\gamma \gamma \ra W^+ W^- \;$}
\newcommand{\gamgamtwwht}{$\gamma \gamma \ra W^+ W^- H \;$}
\newcommand{\gamgamtwwh}{\gamma \gamma \ra W^+ W^- H \;}

\newcommand{\ptu}{p_{1\bot}}
\newcommand{\vecptu}{\vec{p}_{1\bot}}
\newcommand{\ptd}{p_{2\bot}}
\newcommand{\vecptd}{\vec{p}_{2\bot}}
\newcommand{\ie}{{\em i.e.}}
\newcommand{\cm}{{{\cal M}}}
\newcommand{\cl}{{{\cal L}}}
\newcommand{\cd}{{{\cal D}}}
\newcommand{\cv}{{{\cal V}}}
\def\slashc{c\kern -.400em {/}}
\def\slashL{L\kern -.450em {/}}
\def\slashcl{\cl\kern -.600em {/}}
\def\W{{\mbox{\boldmath $W$}}}
\def\B{{\mbox{\boldmath $B$}}}
\def\noi{\noindent}
\def\nn{\noindent}
\def\sm{${\cal{S}} {\cal{M}}\;$}
\def\nph{${\cal{N}} {\cal{P}}\;$}
\def\sb{$ {\cal{S}}  {\cal{B}}\;$}
\def\ssb{${\cal{S}} {\cal{S}}  {\cal{B}}\;$}
\def\cviol{${\cal{C}}\;$}
\def\pviol{${\cal{P}}\;$}
\def\cpviol{${\cal{C}} {\cal{P}}\;$}

\newcommand{\lgg}{\lambda_1\lambda_2}
\newcommand{\lww}{\lambda_3\lambda_4}
\newcommand{\ppin}{ P^+_{12}}
\newcommand{\pmin}{ P^-_{12}}
\newcommand{\ppout}{ P^+_{34}}
\newcommand{\pmout}{ P^-_{34}}
\newcommand{\sinsq}{\sin^2\theta}
\newcommand{\cossq}{\cos^2\theta}
\newcommand{\yt}{y_\theta}
\newcommand{\hppll}{++;00}
\newcommand{\hpmll}{+-;00}
\newcommand{\hpplt}{++;\lambda_30}
\newcommand{\hpmlt}{+-;\lambda_30}
\newcommand{\hpptt}{++;\lambda_3\lambda_4}
\newcommand{\hpmtt}{+-;\lambda_3\lambda_4}
\newcommand{\dk}{\Delta\kappa}
\newcommand{\klam}{\Delta\kappa \lambda_\gamma }
\newcommand{\kac}{\Delta\kappa^2 }
\newcommand{\lac}{\lambda_\gamma^2 }
\def\gamgamtzz{$\gamma \gamma \ra ZZ \;$}
\def\gamgamtww{$\gamma \gamma \ra W^+ W^-\;$}
\def\gamgamtwwe{\gamma \gamma \ra W^+ W^-}

\section{Symmetry breaking and anomalous gauge bosons couplings}
\subsection{The mass Connection}
All data to date, crowned by the results of LEP 1 (dedicated physics with {\em
1} weak
boson) have left no doubt that the Standard Model, \sm, has passed with
flying colours all the low energy tests, even and especially at the quantum
level. Yet, despite the absence of the slightest hint of any anomaly, the model
has still not been elevated to the status of a fully-fledged
theory.
 The reason for this status is essentially due to the sector in
the model that
implements the mass generation and the mechanism of symmetry breaking, \sb.
It is in this sector that originates the remaining missing particle of the
model, the Higgs, about which even the very precise data give no direct
unambiguous clue. Add to this  that an elementary scalar is unnatural, it is no
wonder that almost all the {\em beyond the \sm} activity covered
 by the
various talks at this conference is an investigation or a modification of this
sector. \\
Whatever the structure and the particle content of this sector, we know,
at least,
that it contains

\beqn \label{wmass}
\cl_M=M_W^2 W^+_\mu  W^{-\mu} + \frac{1}{2}M_Z^2 Z_\mu Z^\mu
\eeqn

\noi For the fermionic mass terms our knowledge is even more limited as we do
not
have access to all the elements of the mass matrices of the ups and downs.

\noi The mass term ~(\ref{wmass}) is the most trivial term that may be
regarded as describing
a self-coupling between the $W$'s. These couplings tell another story than
the self-couplings that are present in any unbroken gauge theory like
QCD, say, which originate from the kinetic part of the spin-1 boson and which
describe the propagation and interaction of transverse sates. In the
electroweak model\footnote{The conventions and definitions of the fields and
matrices that I am using here are the same as those in \cite{Hawai}.}

\beqn \label{kinetic}
\cl_G=- \frac{1}{2} \left[
Tr(\W_{\mu \nu} \W^{\mu \nu}) + Tr(\B_{\mu \nu} \B^{\mu \nu}
) \right]
\eeqn

\noi These interactions only  involve the field strength and are thus
explicitely
gauge invariant.  The mass terms, that introduce the longitudinal degrees of
freedom,  would seem to break
this crucial local gauge symmetry. The important point, as you know, is that
the
symmetry is not broken but rather hidden. Upon introducing auxiliary fields
with
the appropriate gauge transformations, we can rewrite the mass term in a
manifestly local gauge invariant way (through the use of covariant
derivatives).
In the minimal standard model this is done
though a doublet of scalars, $\Phi$, of which one is the {\em physical} Higgs.
The
simplest choice of the
doublet implements an extra global {\em custodial} SU(2) symmetry that gives
the well established $\rho=\frac{M_W^2}{M_Z^2 c_W^2} \simeq 1$:

\beqn \label{higgspart}
\cl_{H,M}&=&(\cd_\mu \Phi)^\dagger(\cd^\mu \Phi)\;-\;
\lambda \left[ \Phi^\dagger \Phi - \frac{\mu^2}{2 \lambda}
\right]^2
\eeqn

\noi In the case where the Higgs does not exist or is too heavy, one can modify
this
prescription such that  only the Goldstone Bosons $\omega_{1,2,3}$, grouped in
the matrix $\Sigma$, are eaten (see for instance\cite{Appelquist}):
\beqn \label{masscov}
\cl_M=\frac{v^2}{4} \tr(\cd^\mu \Sigma^\dagger \cd_\mu \Sigma) \;\;\; ; \;\;
\Sigma=exp(\frac{i \omega_{\alpha} \tau^{\alpha}}{v})  \;\;\; (v=246GeV)
\eeqn
In this so-called  non-linear realisation  of \sb
the mass term ~(\ref{wmass}) is formally recovered by going to  the physical
``frame" (gauge)
where all Goldstones disappear, {\it i.e.}, $\Sigma \ra${\bf 1}. \\

\noi The above operators that describe the self-interaction of the vector
bosons
constitute the {\em minimal} set of operators that can be written given the
well-confirmed symmetries of the weak interaction and the known content of the
\sm spectrum. In this sense, the non-linear realisation is even more economical
since it
does not  appeal to the still missing Higgs. These operators are minimal not
only
in the sense of their fields content but also in the sense that these are the
lowest
dimension operators that we can write. In the case of the non-linear
realisation
it is more appropriate to talk about operators with the least number of
derivatives. One expects that, in the absence of a direct observation of new
particles especially those that emerge from the mass sector, phenomena related
to \sb can be described in terms of higher order terms constructed in the mould
of ~(\ref{higgspart},\ref{masscov}). These induce new weak bosons
self-couplings.
To investigate their presence we would then study
interactions involving {\em longitudinal} vector bosons.\\
\noi  Of course, one can
construct other operators describing vector bosons self-couplings  on the mould
of the
{\em universal}
kinetic term ~(\ref{kinetic}) which is explicitely gauge invariant. In this
case
it is worth keeping in mind that these types of anomalies will not be telling
us much about symmetry
breaking, but only that there may be some weakly heavy interacting particles.
The oldest example of such operators for the transverse modes is the celebrated
Euler-Heisenberg Lagrangian that describes  (in the first order)
an {\em anomalous} 4-photon coupling. Given my bias about the importance of
effects intimatley related to \sb I will not be concentrating much on this type
of anomalies, this is the first level where I would like to discriminate
between
origins of anomalies. \\

\noi Another example of an effective Lagrangian that  has proved
more revealing and rich in physics is the
effective chiral  Lagrangian that describes the interaction of pions out of
which one has learnt
so much about the interaction of hadrons. Likewise, one hopes that the
electroweak
equivalent (generalisation of ~(\ref{masscov}) where the pions are to be
identified
with the pseudo-Goldtsone bosons) will teach us something about symmetry
breaking. I will also take the biased point  of view  (level 2 of
discrimination) that
if one still pursues the description of anomalous couplings within the light
Higgs linear approach, then it may be more educating to probe the
characteristics
and the couplings of the Higgs. But this is not the subject of my talk.\\
To summarise at this point, the  type of self-couplings
that, in my view,  deserve
the highest priority are those that one has to probe in the eventuality that
there is no Higgs. This is because I consider that if the Higgs is light one
has
already learnt a great deal about \sb, that the weak interaction will remain
weak at TeV energies and that one should probably concentrate on studying the
spectra of the New Physics that is associated with the symmetry that naturally
accomodates a light Higgs, SUSY.

\subsection{Phenomenological Parameterisation}
The purpose of my rather long  introduction was to stress the connection
between
the investigations of the \sb sector through the study of
anomalous gauge bosons couplings. These would be parameterised by operators of
higher
dimensions  or  of higher order in the energy expansion than those in
(\ref{kinetic}~-~\ref{masscov}). This, of course,
is suggestive of an ordering of operators with respect to the scale of the new
physics (the \sb scale in the point of view I am taking). Of
course, if one is doing experiments at an energy near this scale this ordering
makes no sense and one should consider all the tower of operators. In this
case,
especially for the longitudinal gauge bosons, our underlying gauge symmetry
principle
that is instrumental for the ranking, such as to filter only a small subset of
operators, would not be of much help. Still, we can  use the {\em exact
non-broken} symmetries like the $U(1)_{{\rm QED}}$ and Lorentz invariance
to write all the possible operators that can give an effect to a particular
situation. The {\em phenomenological} parameterisation of the $WW\gamma$ and
$WWZ$ vertex of HPZH
\cite{HPZH} has been written for the purpose of studying \eewwt, the
bread-and-butter of
LEP2. The same parameterisation, although
as general as it can be for \eewwt,  may not be
necessarily correct nor general when applied to other situations. In principle,
if one is guiding by this general principle of keeping only the NON-BROKEN
symmetries,
one should write a new set of operators for every new situation. This does not
necessarily contain all the operators of HPZH. This is one of the shortcomings.
 Nonetheless, the HPZH
parameterisation has become popular enough in discussing anomalies that I will
refer
to it quite often as a common ground when comparing various
approaches and ``data".
To keep the discussion tractable (lack of time) I will only pick out the \cviol
and \pviol
conserving parts of this parameterisation othewise one has to consider in all
generality 13 couplings. Indeed it has been shown\cite{Majorana}
that a particle of spin-J which is not its own anti-particle
can have, at most, $(6J+1)$
electromagnetic form-factors including \cviol, \pviol and \cpviol
violating terms. The
same argument tells us \cite{Majorana}
that if the ``scalar"-part of a massive spin-1
particle does not contribute, as is the case for the Z in
$e^+ e^- \ra W^+ W^-$,
then there is also the same number of invariant form-factors for
the spin-1 coupling to a charged spin-J particle.
The \cviol and \pviol conserving part  of the HPZH\cite{HPZH}  parameterisation
is
\beqn \label{pheno}
{\cal L}_{WWV}&=& -ie \left\{ \left[ A_\mu \left( W^{-\mu \nu} W^{+}_{\nu} -
W^{+\mu \nu} W^{-}_{\nu} \right) \;+\;
\overbrace{ (1+\mbox{\boldmath $\Delta \kappa_\gamma$} )}^{\kappa_\gamma}
F_{\mu \nu} W^{+\mu} W^{-\nu} \right] \right.
\nonumber \\
&+& \left. cotg \theta_w \left[\overbrace{(1+ {\bf \Delta g_1^Z})}^{g_1^Z}
Z_\mu \left( W^{-\mu \nu} W^{+}_{\nu} -
W^{+\mu \nu} W^{-}_{\nu} \right) \;
+ \;
\overbrace{(1+\mbox{\boldmath $\Delta \kappa_Z$} )}^{\kappa_Z}
Z_{\mu \nu} W^{+\mu} W^{-\nu} \right] \right.
\nonumber \\
&+&\left. \frac{1}{M_{W}^{2}}
\left( \mbox{\boldmath $\lambda_\gamma$} \;F^{\nu \lambda}+
\mbox{\boldmath $\lambda_Z$} \;
cotg \theta_w Z^{\nu \lambda}
\right) W^{+}_{\lambda \mu} W^{-\mu}_{\;\;\;\;\;\nu} \right\}
\eeqn

For those not working in the field and who want to get a feeling for what these
form factors mean, suffice it to say that the combination
$\mu_{W}=e (2+\Delta \kappa_\gamma + \lambda_\gamma)/2M_W$ describes
the $W$ magnetic moment and $Q_{W}= - e (1+\Delta \kappa_\gamma
-\lambda_\gamma)/M_W^2$ its quadrupole moment \footnote{The deviations
from the \underline{minimal} gauge value are understood to be evaluated at
$k^2=0$.}.
$(1+ {\bf \Delta g_1^Z})$ can be interpreted as the charge the ``$Z$ sees" in
the
$W$.
Note that the $\lambda$ terms only involve the field strength, therefore they
predominantly affect the production/interaction of transverse $W$'s, in other
words they do not  usefully probe the \sb
sector I am keen to talk about here. \\
Pursuing this observation a little further one can easily describe the
distinctive
effects the other terms have on different reactions and the reason that some
are
found to be
much better constrained in some reactions than others.
First, wherever you look, the $\lambda$'s live in a
world on their own, in the ``transverse world".  If their effect is found to
increase dramatically with energy this is due to the fact that these
are higher order in the energy expansion (many-derivative operators). The other
couplings can also grow with energy if a maximum number of longitudinals are
involved, the latter provide an enhanced strength due to the fact that
the leading term of the longitudinal polarisation is $\propto \sqrt{s}/M_W$.
This enhanced strength does not
originate from the field strength!.
For instance, in \eewwt, $g_1^Z$ produces one $W$ longitudinal
and one transverse: since the produced $W$ come, one from the field strength
the other from the ``4-potential" (longitudinal) whereas the the $\kappa$ terms
produce two longitudinals and will therefore be better constrained in \eewwt.
The situation is reversed in the case of $pp \ra WZ$.  This also tells
us how one may  disentangle between different origins, the reconstruction of
the
$W$ and $Z$
polarisation is crucial. I have illustrated this in
fig.~\ref{lkgfig}, where I have reserved the {\bf thick} arrows for
the ``important' directions:
\begin{figure*}[htb]
\caption{\label{lkgfig}{\em The effect of the phenomenological parameters on
the
vector boson pair production.}}
\begin{center}
\mbox{\epsfxsize=14cm\epsfysize=5cm\epsffile{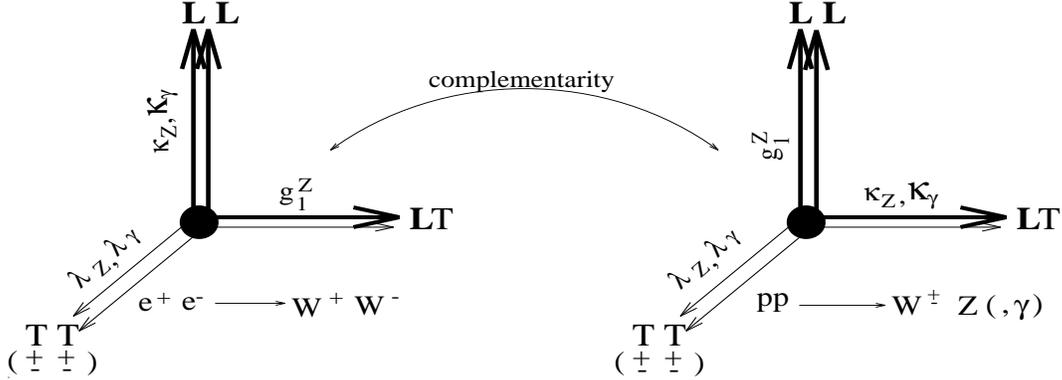}}
\vspace*{-1cm}
\end{center}
\end{figure*}
\vspace*{0.5cm}

There are some limits on these couplings from CDF/D0\cite{CDF} extracted from
the study
of $WZ, WW$ and $W\gamma$ production:
$-2.3 < \Delta \kappa_\gamma <2.2 \;;\; -0.7 < \lambda_\gamma < 0.7\;$
while a constrained global fit with  $\lambda_\gamma=\lambda_Z$,
$\kappa_\gamma=\kappa_Z (g_1^Z=1)$ gives
$-0.9 < \Delta \kappa_V <1. \;;\;-0.5 < \lambda_V < 0.5$.\\
\noi I would like to argue that these values are too large to be meaningful.
These
are too large in the sense that they can hardly be considered as precision
measurements, a far cry from the precision that one has obtained on the
vector-fermion
couplings at LEP1! In the case of the Tevatron and $W$
self-couplings one is talking about deviations of order $100\%$!

\noi Talking about the LEP1 data, with this year's statistics  one is now
sensitive to
the genuine non-Abelian radiative corrections and therefore to the presence of
the tri-linear (and quadrilinear) couplings \cite{Gambino}. Even so,
the data gives no clear information about the presence of the Higgs. In my view
this should be taken as  very strong evidence for the \su local gauge symmetry
or more precisely that the higher order terms that may correct (\ref{kinetic})
must {\em naturally} be small. On the other hand the \sb sector apart from the
mass terms still keeps its secret.

\noi It is worth stressing again, contrary to the fierce attack
\cite{Ruj}
that the above HPZH
Lagrangian (eqt.~(\ref{pheno})) is not locally gauge invariant and leads to
trouble at the quantum
level, that as the lenghty introduction has shown all the above operators can
be
made gauge invariant, by unravelling and making explicit  the compensating
Goldstone fields and
extra vertices that go with the above. Under this light, the HPZH
parametrisation
should be considered as being written in a specific gauge and that after this
gauge (unitary) has been chosen it is non-sensical to speak of gauge
invariance\cite{Cliff}.
But of course, it is much much better to keep the full symmetry so that one can
apply the Lagrangian to any situation and in any frame.  There is another
benefit in doing so. If the scale of new physics is far enough compared to the
typical energy where the experiment is being carried out\footnote{If this is
not the case then we should see new particles or at  least detect their
tails.},
then one should only include the first operators in the energy expansion,
beyond
those of the \sm. Doing so will maintain some constraints on the parameters
$\lambda, g_1^Z, \Delta \kappa$. These constraints will of course be lost if
you
allow higher and higher order operators or allow strong breaking of custodial
symmetry, in both cases rendering the situation chaotic while LEP1 shows and
incredible regularity. It is highly improbable that the order and symmetry is
perturbed so badly.

So what are these operators that describe the self-couplings when one restricts
one-self to next-to-leading operators by exploiting the \su and the custodial
symmetry? and how are they mapped on the HPZH phenomenological parameters?
These are given in Table~1. for the linear \cite{BuchWy,Ruj} as well as the
non-linear realisation\cite{Hawai,Holdom} to
bring out some distinctive features about the two approaches:
\begin{table*}[htbp]
\caption{\label{linearvsnonlinear}
{\em The Next-to-leading Operators describing the $W$ Self-Interactions which
do not contribute to the $2$-point function.}}
\vspace*{0.3cm}
\centering
\begin{tabular}{|l||l|}
\hline
{\bf Linear Realization \hspace*{0.1cm}, \hspace*{0.1cm} Light Higgs}&
{\bf Non Linear-Realization \hspace*{0.1cm}, \hspace*{0.1cm} No Higgs}\\
\hline
&\\
${{\cal L}}_{B}=i g' \frac{\epsilon_B}{\Lambda^2} (\cd_{\mu}
\Phi)^{\dagger} B^{\mu \nu} \cd_{\nu} \Phi$&
${{\cal L}}_{9R}=-i g' \frac{L_{9R}}{16 \pi^2} \tr (  \B^{\mu \nu}\cd_{\mu}
\Sigma^{\dagger} \cd_{\nu} \Sigma )$ \\
&\\
${{\cal L}}_{W}=i g \frac{\epsilon_w}{\Lambda^2} (\cd_{\mu}
\Phi)^{\dagger} (2 \times \W^{\mu \nu}) (\cd_{\nu} \Phi)$&
${{\cal L}}_{9L}=-i g \frac{L_{9L}}{16 \pi^2} \tr ( \W^{\mu \nu}\cd_{\mu}
\Sigma \cd_{\nu} \Sigma^{\dagger} ) $ \\
&\\
$\cl_{\lambda} = \frac{2 i}{3} \frac{L_\lambda}{ \Lambda^2}
g^3 \tr ( \W_{\mu \nu} \W^{\nu \rho} \W^{\mu}_{\;\;\rho})$&
$\;\;\;\;\;\;\;\;---------\;\;\;\;$\\
&\\
$\;\;\;\;\;\;\;\;---------\;\;\;\;$&
$\cl_{1}=\frac{L_1}{16 \pi^2} \left( \tr (D^\mu \Sigma^\dagger D_\mu \Sigma)
\right)^2\equiv \frac{L_1}{16 \pi^2} {{\cal O}}_1$ \\
$\;\;\;\;\;\;\;\;---------\;\;\;\;$&$
\cl_{2}=\frac{L_2}{16 \pi^2} \left( \tr (D^\mu \Sigma^\dagger D_\nu \Sigma)
\right)^2 \equiv \frac{L_2}{16 \pi^2} {{\cal O}}_2$ \\
& \\
\hline
\end{tabular}
\end{table*}

\noi By going to the
physical gauge, one recovers the phenomenological parameters with the
{\em constraints}:
\beqn \label{constraints}
\Delta\kappa_\gamma&=&
\frac{e^2}{s_w^2} \frac{v^2}{4 \Lambda^2}
(\epsilon_W+\epsilon_B )=
\frac{e^2}{s_w^2} \frac{1}{32 \pi^2} \left( L_{9L}+L_{9R} \right)
\nonumber \\
\Delta\kappa_Z&=&\frac{e^2}{s_w^2} \frac{v^2}{4 \Lambda^2}
(\epsilon_W -\frac{s_w^2}{c_w^2} \epsilon_B) =
\frac{e^2}{s_w^2} \frac{1}{32 \pi^2} \left( L_{9L}
-\frac{s_w^2}{c_w^2} L_{9R} \right)
\nonumber \\
\Delta g_1^Z&=&\frac{e^2}{s_w^2} \frac{v^2}{4 \Lambda^2}
(\frac{\epsilon_W}{c_w^2})=
\frac{e^2}{s_w^2} \frac{1}{32 \pi^2} \left(\frac{ L_{9L}}{c_w^2} \right)
\nonumber \\
\lambda_\gamma&=&\lambda_Z=\left(\frac{e^2}{s_w^2}\right)
L_\lambda \frac{M_W^2}{\Lambda^2}
\eeqn

 {\em Catch 22:}\\
Aren't there other operators with the same symmetries that appear at the {\em
same
level} in the hierarchy and would therefore be as likely? \\
\noi \underline{Answer:} YES. And this is an upsetting conceptual problem.
On the basis of the above symmetries, one can not help it, but
there are other operators which contribute to
the tri-linear couplings and have a part which corresponds to
bi-linear anomalous
$W$ self-couplings. Because of the latter and of the unsurpassed precision
of LEP1, these operators are already very much {\em unambiguously}
constrained.  Examples of
such annoying operators in the two approaches are
\beqn
{{\cal L}}_{WB}&=&g g' \frac{\epsilon_{WB}}{\Lambda^2}\; \left(
\Phi^{\dagger} \times \W^{\mu \nu} \Phi \right)  B_{\mu \nu}
\nonumber \\
{{\cal L}}_{10}&=&g g' \frac{L_{10}}{16 \pi^2} \tr ( \B^{\mu \nu}
\Sigma^{\dagger} \W^{\mu \nu}  \Sigma ) \longrightarrow L_{10}=-\pi S\simeq
\frac{4 \pi
s_W}{\alpha} \epsilon_3
\eeqn
For example, current limits from LEP1 indicate that $-1.4<L_{10}<2.$ This is
really small, so small that if the other $L_i$'s, say, were of this order it
would be
extremely difficult to see any effect at the next colliders. So why should the
still not-yet-tested operators be  much larger? This is the naturalness
argument
which in my view is {\em the} essential point of \cite{Ruj}.
One can try hard to find models with no contributions to $L_{10}$. But the
solutions are either not very appealling or one has to accept that this
quantity
can not be calculated reliably in the context of non-perturbative models
(in the non-linear approach). Of course, there is also the easy escape
that we have not been ingenious enough.... \\
 To continue with my talk I will assume
that $L_{10}\sim 0$ can be neglected compared to the other operators.
This said, I will not completely ignore this limit and the message that LEP1 is
giving us, especially that various arguments about the ``natural order of
magnitude"\footnote{Refer to the talk of Wudka.}
for these operators should force one to consider a limit extracted from future
experiment to be meaningful if $|L_i|<\sim 10$ (10 is really generous..).
This translates into
$\Delta\kappa, \dgz <\sim 10^{-2}$. Note that the present Tevatron limits if
they were to be written in terms of $L_{9}$ give $L_9\sim 10^3$!!!.

\noi With this caveat about $L_{10}$ and the like, let us see how the 2
opposite
assumptions about the lightness of the Higgs differ in their most probable
effect on the $W$ self-couplings. First, the tri-linear coupling $\lambda$ is
relegated to higher orders in the heavy Higgs limit(less likely).
This is as expected: transverse
modes are not really an issue here. The main difference is that with a heavy
Higgs, genuine quartic couplings contained in $L_{1,2}$ are as likely as the
tri-linear
 and, in fact, when
contributing to $WW$ scattering their effect will by far exceed that of the
tri-linear.
This is because $L_{1,2}$ involve essentially longitudinals. This is another
way
of arguing that either the Higgs exists or expect to ``see something" in $WW$
scattering. Note also that $L_{9L,W,\lambda}$ do give quadri-linear bits but
these are imposed by gauge invariance. Note also that $L_{9R, B}$ is not
expected
to contribute significantly in $pp \ra WZ$ since it has no contribution to
$\Delta g_1^Z$ (see fig.~1). This is confirmed by many analyses.


Going to the physical gauge, the quartic couplings from the chiral
approach are
\beqn
\cl^{SM}_{WWV_{1}V_{2}} &=& -e^2 \left\{ \left(
A_\mu A^\mu W^{+}_{\nu} W^{- \nu} - A^\mu A^\nu W^{+}_{\mu}
W^{-}_{\nu} \right) \right. \nonumber \\
&+& 2 \frac{c_w}{s_w} (1+\frac{l_{9l}}{c_w^2}) \left(
A_\mu Z^\mu W^{+}_{\nu}W^{-\nu} - \frac{1}{2}
A^\mu Z^\nu ( W^{+}_{\mu}W^{-}_{\nu} + W^{+}_{\nu}W^{-}_{\mu} ) \right)
\nonumber \\
&+&\frac{c_w^2}{s_w^2} (1+\frac{2 l_{9l}}{c_w^2}-\frac{l_-}{c_w^4}) \left(
Z_\mu Z^\mu W^{+}_{\nu}W^{-\nu} - Z^\mu Z^\nu W^{+}_{\mu}W^{-}_{\nu} \right)
\nonumber \\
&+& \frac{1}{2 s_w^2} (1+2 l_{9l}-l_-)
 \left(W^{+\mu} W^{-}_{\mu} W^{+\nu} W^{-}_{\nu} -
W^{+ \mu} W^{+}_{\mu} W^{-\nu}W^{-}_{\nu} \right) \nonumber \\
&-&\frac{l_+}{2s_w^2}\left( \left( 3 W^{+\mu} W^{-}_{\mu} W^{+\nu} W^{-}_{\nu}+
W^{+ \mu} W^{+}_{\mu} W^{-\nu}W^{-}_{\nu} \right) \right. \nonumber \\
&+&\frac{2}{c_w^2} \left. \left. \left(Z_\mu Z^\mu W^{+}_{\nu}W^{-\nu} + Z^\mu
Z^\nu W^{+}_{\mu}W^{-}_{\nu}
 \right) + \frac{1}{c_w^4} Z_\mu Z^\mu  Z_\nu Z^\nu \right)\right\} \nonumber
\\
&\;& {\rm with}\;\;\;\;\;\;l_{9l}=\frac{e^2}{32 \pi^2 s_w^2} L_{9L}\;\;\;
;\;\;\;l_\pm=\frac{e^2}{32 \pi^2 s_w^2}
(L_1 \pm L_2)
\eeqn
Note that the genuine trilinear $L_{9L}$ gives structures analogous to the \sm.
The two photon couplings (at this order) are untouched by anomalies.

\section{Future Experimental Tests}
With the order of magnitude on the $L_i$ that I have set as a meaningful
benchmark,
one should realise that to extract such
(likely) small numbers one needs to know the \sm cross sections with a
precision
of the order of $1\%$ or better. This calls for the need to include the
radiative
corrections especially the initial state radiation. Moreover one should try to
extract as much information from the $W$ and $Z$ samples: reconstruct the
helicities, the angular distributions and correlations of the decay products.
These criteria mean precision measurements and therefore we expect \epemt
machines to have a clear advantage assuming that they have enough energy.
Nonetheless, it is instructive to refer to fig.~\ref{lkgfig} to see that $pp$
machines
could be complementary.\\
\begin{figure*}[htbp]
\caption{\label{l9fig}{\em Comparison between the expected bounds on the
two-parameter
space $(L_{9L},L_{9R}) \equiv (L_W,L_B) \equiv (\Delta g_1^Z,
\Delta \kappa_\gamma) $ (see text for
the conversions) at the NLC500 (with no initial polarisation), LHC
and LEP2. The NLC bounds are from $e^+e^- \ra W^+W^-\;,W^+W^-\gamma, W^+W^-Z$
(for the latter these are one-parameter fits) and $\gamma \gamma \ra W^+W^-$.
The LHC bounds are from $pp \ra WZ$.
Limits from a single parameter fit are also shown (``bars") .}}
\begin{center}
\vspace*{-2.5cm}
\mbox{\epsfxsize=14.5cm\epsfysize=21cm\epsffile{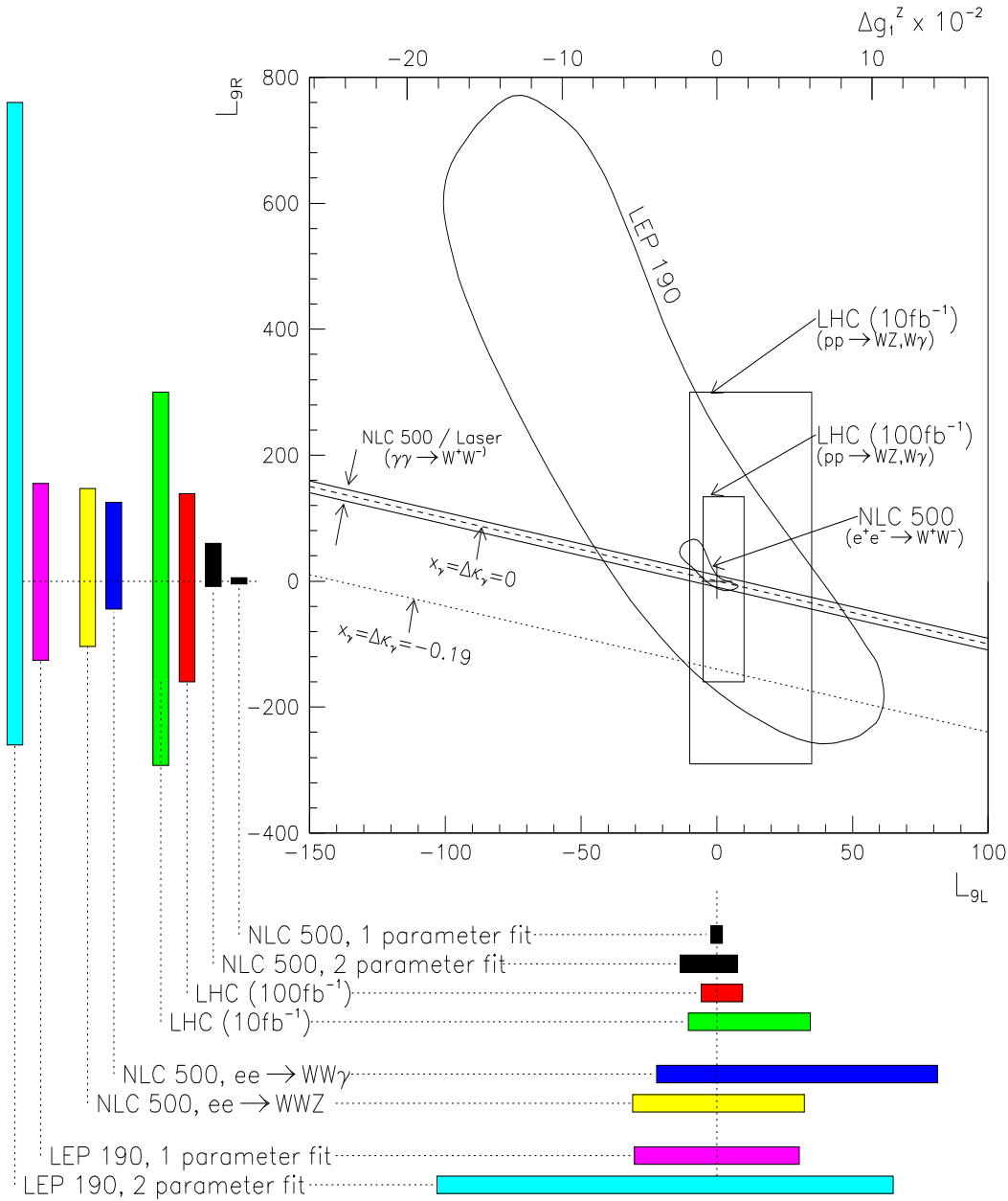}}
\vspace*{-2.5cm}
\end{center}
\end{figure*}
In the following, one should keep in mind that all the extracted limits fall
well
within the unitarity limits. I only discuss the description in terms of
``anomalous couplings" below an effective cms energy of a VV system  $\sim
4 \pi v \sim 3-4 TeV$, without the inclusions of resonances\footnote{At this
conference, this  has been discussed by  Roberto Casalbuoni and  Kingman
Cheung.}.
Moreover, I will not discuss the situation when parameters are dressed with
energy dependent form factors or any other scheme of unitarisation that
introduces more model dependence on the extraction of the limits.
For reasons of space I will not go into the details of how the various
operators
are looked for in various processes and different machines ($pp, \epem,
\gamma\gamma$) but refer to  a summary  I have given
elsewhere\cite{Hawai}. I will, however, update some of the results and
summarise them in the comparative figure that gives the  limits on the genuine
tri-linear
couplings. These limits are given in terms of the chiral Lagrangian parameters
$L_{9L,R}$ or equivalently using (\ref{constraints})  in terms of $L_{B,W}$.
They can also be
re-interpreted in terms of the more usual $\kappa_V,g_1^Z$ with the constraint
given
by (\ref{constraints}), in which case the $L_{9L}$ axis is directly
proportional
to \dgzt. \\

The limits from $pp$ that I have given in\cite{Hawai} are obsolete
(in the present updated version $pp$  means LHC with two settings for the
luminosity $10$ and $100fb^{-1}$). The new limits are based on a very careful
study\cite{Baur} that includes the very important effect of the QCD
NLO corrections as well as implemeting the full spin correlations for the most
interesting channel $pp \ra WZ$. $WW$ production
with $W\ra jets$ production is fraught with a huge QCD background,  while the
leptonic mode is extremely difficult to reconstruct due to the 2 missing
neutrinos. The NLO corrections for $WZ$ production are huge, especially
in   precisely the regions
where the anomalous are expected to show up\cite{Baur}. For instance, high
$p_T^Z$.
In the inclusive cross section this is mainly due to, first, the importance
of the subprocess $q_1 g\ra Z q_1$ (large gluon density at the LHC) followed by
the ``splitting" of the quark $q_1$ into $W$. The probability for this
splitting
increases with the $p_T$ of the quark (or Z): Prob$(q_1\ra q_2 W) \sim
\alpha_w/4\pi ln^2(p_T^2/M_w^2)$. To reduce this effect one \cite{Baur} has to
define
an exclusive cross section that should be as close to the LO $WZ$ cross section
as possible by cutting on the extra high $p_T$ quark (dismiss any jet with
 $p_T^{{\rm jet}}>50GeV, |\eta_{{\rm jet}}|<3$). This defines a NLO
$WZ +``0{{\rm jet"}}$ cross section which
is stable against variations in the choice of the $Q^2$ but which nonetheless
can
be off by as much as $20\%$ from the prediction of Born \sm result.
The anomalous parameters are included one by  one in the form of the HPZH
parameterisation. It is indeed found, as expected from the general arguments
that
I exposed above, that \dgzt is much better constrained than $\Delta \kappa_Z$.
I have thus reinterpreted the results in the chiral Lagrangian approach
approximating the effect of $L_{9L}$ as being dominantly due to \dgzt  while
I blamed the bad limit on $\Delta \kappa_Z$ on $L_{9R}$.

For the case of \epemt at high energies, the comparative figure shows the
adaptation of the  {\em BM2}\cite{BM2} results. These are based on a very
powerful fitting
procedure that aims at reconstructing {\bf 8} observables which are
combinations of density matrices. Simulations performed for LEP2
energies by
experimentalists\cite{Sekulin} have shown that we can somehow improve on these
limits. The main missing ingredient that may change these results is, once
again, the effect of radiative corrections. Notably, bremstrahlung and
beamstrahlung were not taken into account. It is now mandatory to include these
corrections, for a review see\cite{Wim}.
As in the case of $pp$, initial state radiation drastically affects some of the
distributions that,
at tree-level, seem to be good New Physics discriminators. For instance,
initial
state radiation is responsible for the boost effect that redistributes phase
space: this leads to the  migration of the forward $W$ into the backward
region and results in a large correction in the backward region. Precisely the
region where one would have hoped to see any s-channel effect more clearly.
Second,  if one reconstructs the polarisation of the $W$
without taking into account the energy loss,  one may ``mistag"
a transverse W for a  longitudinal, thereby introducing a huge correction in
the small
tree-level longitudinal cross sections, which again is
 particularly sensitive to New Physics.
Cuts must be included. With the near advent of LEP2, there is
 now the discussion\cite{Excalibur} whether
an analysis based on the resonant diagrams is enough. It is found that non
resonating (non genuinely $WW$) 4-fermion states are not negligible. Probably,
it
is best to cut on the non-resonant diagrams  by double mass constraints (etc..)
at the expense of reducing the event sample, rather than working with a
``mixed" final
state.

\begin{figure*}[htbp]
\caption{\label{barklow.fig}{\em Limits on ($L_{9L}-L_{9R}$) in
\epemt including ISR and beam polarisation.}}
\begin{center}
\mbox{\epsfxsize=8cm\epsfysize=8cm\epsffile{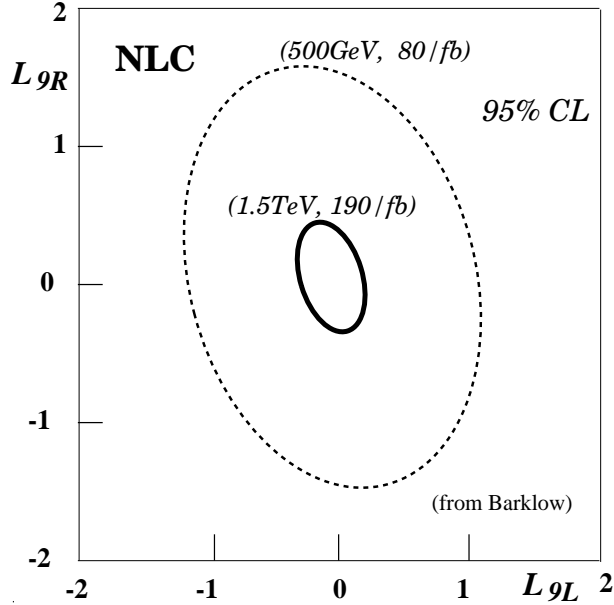}}
\vspace*{-1cm}
\end{center}
\end{figure*}
\vspace*{0.5cm}

\noi Very recently Barklow\cite{Barklow} has reanalysed the operators
$L_{9L,9R}$ by considering the correlated 4-fermion-$WW$ five-fold angular
distributions and including NLC luminosity spectra as well as considering the
effect
 of initial
polarisation. The latter, as is known, can easily isolate the the s-channel
$WWV$.
 His analysis at $500$GeV assumes a luminosity of $80fb^{-1}$, which is much
larger than
what has been assumed in the similar study of {\em BM2} ($10fb^{-1}$).
However, since the sensitivity to the anomalous goes like  $\sim \sqrt{{\cal
L}}$
this confirms the {\em BM2} results and hints that although the inclusion of
the
luminosity spectra makes the analysis more complicated it, fortunately, does
not
critically degrade the sensitivity to the anomalous. Anyway, with this
luminosity the results are fascinating, one can be sensitive to values as
low as 1-2 for the parameters $L_9$. This is really precision measurement.
Moreover, in future \epemt
linacs one also hopes to have a \gamgamt version. A new analysis\cite{Parisgg}
shows
that in combination with the \epemt mode, the \gamgamt mode can help  put
much stronger limits on the parameter space of the anomalous (see fig.~2).\\

In conclusion, it is clear that already with a $500$GeV \epemt collider
combined with a
good integrated luminosity of about $80fb^{-1}$ one can reach a precision,
on the parameters that probe \sb in the genuine tri-linear $WWV$ couplings, of
the same order as what we can be achieved with LEP1 on the two-point vertices.
To
reach higher precision and critically probe \sb one needs to go to $TeV$
machines, as fig.~3 shows for the tri-linear $L_{9L}-L_{9R}$. In fact,  at an
effective $WW$ invariant masses of order the TeV,
\sb (especially in scalar-dominated models) is best probed through the genuine
quartic
couplings in $WW$ scattering or even perhaps in $WWZ, ZZZ$ production (that are
poorly
constrained at $500$GeV). LHC could also address this particular issue but one
needs dedicated careful simulations to see whether any signal could be
extracted in the $pp$ environment. In this regime
there is also the fascinating aspect of $W$ interaction that I have not
discussed
and which is the appearance of strong resonances. This would reveal another
alternative to the \sm description of the scalar sector. \\

\noi {\em {\bf Acknowledgments:}}\\
\noi It is a pleasure to thank Marc Baillargeon, Genevi\`eve B\'elanger,
 Frank Cuypers, Norman
Dombey, and
Ilya Ginzburg for the enjoyable collaborations and discussions. I thank
Misha Bilenky for providing the data for the $L_9$ fits in \epemt. I also thank
the organisers for their kind invitation. \\

\end{document}